\documentclass[twocolumn,showpacs,preprintnumbers]{revtex4} 

\input{epsf}

\def\gev{\mbox{GeV}}

\def\AJ{{\it Ap. J.} }

\def\MNRAS{{\it Mon. Not. R. Ast. Soc.} }

\def\NP{{\it Nucl. Phys.} }
\def\PL{{\it Phys. Lett.} }
\def\PR{{\it Phys. Rev.} }
\def\PRL{{\it Phys. Rev. Lett.} }

\def\frac#1#2{{\textstyle{{#1}\over {#2}}}}

\def\lsim{\mathrel{\rlap{\lower4pt\hbox{\hskip1pt$\sim$}}
    \raise1pt\hbox{$<$}}}
\def\gsim{\mathrel{\rlap{\lower4pt\hbox{\hskip1pt$\sim$}}
    \raise1pt\hbox{$>$}}}
\def\sqr#1#2{{\vcenter{\vbox{\hrule height.#2pt
         \hbox{\vrule width.#2pt height#1pt \kern#1pt
         \vrule width.#2pt}
         \hrule height.#2pt}}}}

\def\beq{\begin{equation}}
\def\eeq{\end{equation}}
\def\beqa{\begin{eqnarray}}
\def\eeqa{\end{eqnarray}}

\begin{document}

\title{WMAP and  Supergravity Inflationary Models}

\author{M. C. Bento}

\altaffiliation[Also at ] {CFIF, Instituto Superior T\'ecnico, Lisboa.
  Email address: bento@sirius.ist.utl.pt}

\author{N. M. C. Santos}

\altaffiliation[Also at ] {CFIF, Instituto Superior T\'ecnico, Lisboa.
 Email address: ncsantos@gtae3.ist.utl.pt }

\author{A. A. Sen}

\altaffiliation[Also at ] {CENTRA, Instituto Superior T\'ecnico, Lisboa.
 Email address: anjan@x9.ist.utl.pt}

\affiliation{ Departamento de F\'\i sica, Instituto Superior T\'ecnico \\
Av. Rovisco Pais 1, 1049-001 Lisboa, Portugal}

\vskip 0.5cm

\date{\today}

\begin{abstract}

We study a class of $N=1$ Supergravity inflationary models in which
the evolution of the inflaton dynamics is controlled by a single power 
in the inflaton field  at the point where the observed density
fluctuations are produced, in the context of the braneworld scenario,
in light of WMAP results. In particular, we find that the bounds on the
spectral index and its running constrain the parameter space  both for
models where the inflationary potential is dominated by a quadratic
term and by a cubic term in the inflaton field. We also find that
$\alpha_s>0$ is required for the quadratic model whereas $\alpha_s<0 $
for the cubic model. Moreover, we have determined an  upper bound on
the five-dimensional Planck scale, $M_5 \lsim 0.019$ M, for the
quadratic model. On the other hand, a running spectral index with
$n_s>1$ on large scales and $n_s<1$ on small scales is not possible
in either case.
 
\vskip 0.5cm
 
\end{abstract}


\maketitle
\section{Introduction}

The first year WMAP data  has confirmed the ``concordance'' values of the
cosmological parameters with unprecedented accuracy
and given important information on the primordial spectrum of 
density perturbations \cite{Bennet,Spergel}. Their results favor
 gaussian, purely adiabatic
fluctuations and a spectral index that runs from $n_s>1$ on large
scales to $n_s<1$ on small scales. Moreover, WMAP has confirmed
earlier COBE DMR observations that there is a lower amount of power on
the largest scales when compared to that predicted by the standard
$\Lambda$CDM models.

 Although these results are not yet firmly established (for an analysis of
 WMAP results which finds no evidence of running see
 Ref.~\cite{BargerSeljak}), 
it seems worthwhile to reexamine inflationary models in light of WMAP results,
 as they may give us further insight into the very early universe. 
Supergravity inflationary models are particularly important as  
supersymmetry (or its local version, supergravity) is the only known
 way to avoid the hierachy problem,
 i.e. the fact  that the high energy scale of inflation communicates to other
 sectors
of the theory driving the electroweak scale much above its  observed
 value   via radiative corrections.

However, supergravity inflationary models also suffer from a kind of hierarchy
 problem as  supersymmetry is broken by the large cosmological
 constant during inflation 
 giving all scalars, including the inflaton, 
 a soft mass of the order of the Hubble
 parameter \cite{Bertolami}. As a result, the curvature of the
 inflaton potential, as
 measured
by the $\eta$ slow-roll parameter, becomes
 too large to allow for a sufficiently long period of inflation to
 take place  - the so-called $\eta$ problem. 

Recently, it has been shown that this problem can be avoided within
 the Randall-Sundrum Type II braneworld scenario \cite{Randall}, at least
 for a class of
supergravity models in which  the evolution 
of the inflaton dynamics is controlled by a single power at the point 
where the observed density fluctuations are produced and   
the inflationary potential can, therefore,  be approximately given by

\beq
V \simeq \Delta^{4} \left[1 + c_n \left(\phi\over M\right)^n\right]~~,
\label{eq:gpot}
\eeq
where $M=M_P/\sqrt{8 \pi}$ is the reduced Planck mass. In 
 the braneworld context, 
 the Friedmann equation acquires an additional term quadratic in
the energy density \cite{Binetruy}

\beq
H^2 = {8 \pi \over 3 M_P^2} \rho \left[1 + {\rho \over 2 \lambda}\right]~,
\label{eq:H22}
\eeq
where $\lambda$ is the brane tension, which relates
the four and five-dimensional Planck scales through

\beq
M_P = \sqrt{{3 \over 4 \pi}} {M_5^3 \over \sqrt{\lambda}}~~.
\label{eq:MP}
\eeq
It is precisely the  new parameter $M_5$ that plays a crucial role
 in the resolution of
 the $\eta$-problem in supergravity inflation. As shown in
  Ref.~\cite{Bento2}, for the  case
 where the first term in Eq.~(\ref{eq:gpot}) is dominant and
$n=2$ or  $n=3$, 
 this problem can be avoided 
provided the five-dimensional  Planck mass satisfies, respectively,
  the condition 
$M_5 \lsim 10^{16}~\gev$ and $M_5 \simeq 1.1 \times 10^{16}~\gev$. 
The case where the second term is dominant and $n=2$, corresponding 
to chaotic inflation, has 
been studied  in Refs. \cite{Maartens,Bento1}, where it is shown that
it is possible to achieve 
successful inflation with sub-Planckian field  values, thereby
avoiding
well known difficulties with higher order non-renormalizable
terms.

In this paper, we reexamine this class of supergravity models for the quadratic
 and cubic cases in light of WMAP results. The case of chaotic
 inflation has already been analysed in Ref.~\cite{Liddle1}, with the
 conclusion that the quadratic potential is allowed at two-sigma for
 any value of the brane tension and the quartic potential is very
 constrained, particularly in the case where the inflationary energy
 scale is close to the brane tension. 
 Here we concentrate on the case where the first term in Eq.~
(\ref{eq:gpot}) is
 dominant. We find that although the running
of the scalar spectral index is within the bounds determined by WMAP
  for  both  the quadratic and cubic models
 and for a
wide range of potential parameters, a  spectral index running from 
 $n_s>1$ on large scales to  $n_s<1$ on small scales is not possible
in either case as precisely the opposite trend is found.

\section{Quadratic potential}

We first consider the case where the potential is 
quadratic in the inflaton field, $\phi$,  and we rewrite it as

\beq
V= V_0  + {1\over 2} m^2 \phi^2~~,
\label{eq:V2}
\eeq
and assume that the first term is dominant.

In supergravity, effective mass squared contributions of fields are given by

\beq
{1\over 2} m^2= 8 \pi {V_0\over M_P^2} \approx 3 H^2~,
\label{scond}
\eeq
since the horizon of the inflationary De Sitter phase has a Hawking 
temperature given by $T_H = H/2\pi$ \cite{Bertolami}. 

Contributions like the ones of Eq. (\ref{scond}) lead to 
$\eta \equiv M_P^2 V^{\prime\prime}/8 \pi V \simeq 2$ ; however, the onset of 
inflation requires $\eta \ll 1$. Within the braneworld scenario, however, 
$\eta$ and  the remaining ``slow-roll'' parameters
 $\epsilon$ and $\xi$,
 are  modified, at high energies, by a factor proportional to 
$\lambda/V$  \cite{Maartens}

\begin{eqnarray}
\label{eq:epsilon}
\epsilon &\equiv&{M_{P}^2 \over 16\pi} \left( {V' \over V}
\right)^2  {1+{V/ \lambda}\over(1+{V/2\lambda})^2}~~,\\
\label{eq:eta}
\eta &\equiv& {M_{P}^2 \over 8\pi} \left(
{V'' \over V} \right)  {1 \over 1+{V/ 2 \lambda}}~~,\\
\xi & \equiv & {M_P^4\over (8 \pi)^2}{V^\prime V^{\prime\prime\prime}\over V^2}
{1\over (1+V/2 \lambda)^2}.
\label{eq:xi}
\end{eqnarray}
In the high energy approximation, $V\gg \lambda$,  we obtain, for this model

\begin{eqnarray}
\epsilon &= & {M_P^2\over 8 \pi}{m^4\phi^2\over V_0^2\alpha}~,\\
\eta &= &{4\over{\alpha}}~,\\
\xi &=&  0   ~,
\end{eqnarray}
where we have also used the approximation $V\simeq V_0$ during
 inflation  and the definition
  $\alpha \equiv V_0/ \lambda$.
As shown in Ref.~\cite{Bento2}, if $\alpha$ is sufficiently large,
 the $\eta$-problem 
is automatically solved by the brane correction.

The number of $e$-folds during inflation, $N$, in the braneworld
scenario, is given by \cite{Maartens}

\beq
\label{eq:N}
N \simeq - {8\pi  \over M_{P}^2}\int_{\phi_{I}}^{\phi_{
F}}{V\over V'} \left[ 1+{V \over 2\lambda}\right]  d\phi~~,
\eeq
in the slow-roll approximation. We see that, as a result of the modification
in the Friedmann equation,
the expansion rate is increased, at high energies,  by a factor $V/2\lambda$.
For this model, we get, in the high-energy approximation,

\beqa
N =&&  {\alpha\over 4} \log\left({\phi_I\over \phi_F}\right)
+ {2 \pi \alpha \over M_P^2}(\phi_I^2-\phi_F^2)\nonumber\\
&&+ {16 \pi^2 \alpha\over M_P^4}(\phi_I^4-\phi_F^4)~.
\label{eq:N2}
\eeqa
Notice that, for sub-Planckian field values, the second and third
terms are negligible.
The value of $\phi$ at the end of inflation can be obtained 
from the condition

\beq
{\rm max}\{\epsilon(\phi_F),|\eta(\phi_F)|\}= 1~~.
\label{eq:phif}
\eeq

However, we shall consider 
$\phi_F=\beta M$ as a free parameter since  quadratic 
potentials of the type we are studying arise typically  in the context of 
hybrid inflation, once some other field is held 
at the origin by its interaction with $\phi$.  In these scenarios,
inflation may end due to instabilities 
triggered by the dynamics of the other field and, therefore, the amount 
of inflation strongly depends on the value of the inflaton field at
the end of inflation, 
$\phi_F$,  at the time the instabilities arise. 
Actually,  these 
instabilities are  necessary in order to end inflation as 
$\epsilon \ll 1$ for $\alpha \gg 1$  
and sub-Planckian field values. 
\begin{figure*}[ht!]
\centering
\leavevmode \epsfysize=14cm \epsfbox{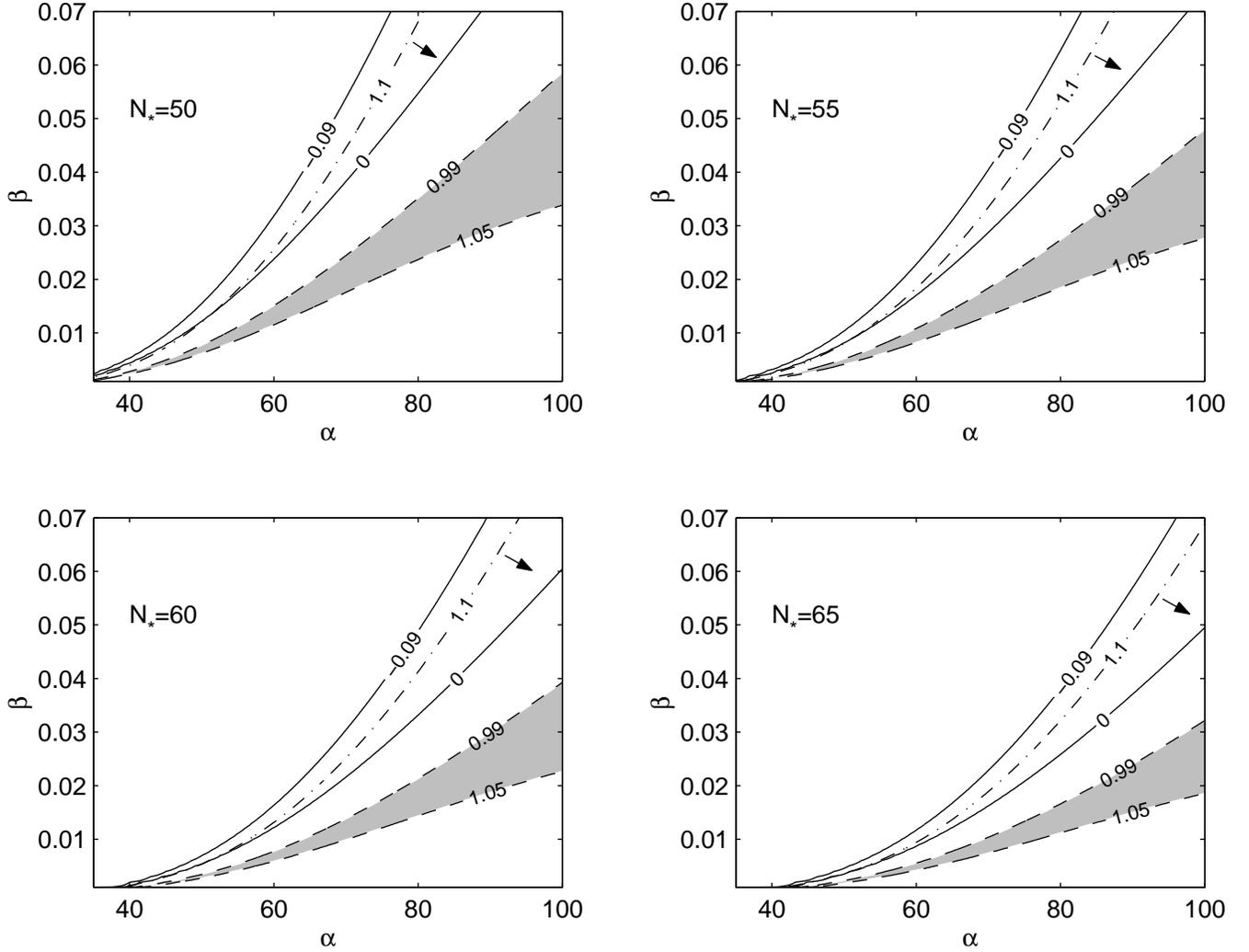}
\vskip 0.1cm
\caption{Contours of the inflationary observables  $n_s$ (dashed), $\alpha_s$
(full) and $r_s$ (dot-dashed) in the ($\alpha, \beta$) plane for the
 quadratic model; the arrow points the direction of decreasing $r_s$.
 The allowed region is shaded and, as
 indicated, 
different panels
 correspond to different values of $N_\star$.}
\label{fig:figure1}
\end{figure*}

The amplitude of scalar perturbations is given by \cite{Maartens}

\beq
\label{eq:As}
A_{s}^2 \simeq \left . \left({512\pi\over 75 M_P^6}\right) {V^3
\over V^{\prime2}}\left[ 1 + {V \over 2\lambda} \right]^3
\right|_{k=aH}~~,
\eeq
where the right-hand side should be evaluated as the comoving wavenumber
equals the Hubble radius during inflation, $k=a H$. Thus the amplitude
of scalar perturbations is increased
relative to the standard result at a fixed value of $\phi$ for a given
potential.
Using the high energy approximation and $V \simeq V_0$ in 
Eq.~(\ref{eq:As}), we obtain 

\beq
A_s^2 \simeq {1600 \pi\over 75}{V_0^6\over \lambda^3 m^4}\exp\left(-8
N_k\over \alpha\right)~~,
\label{eq:As2}
\eeq
where $N_k$ is the number of e-folds between the time the scales of
interest leave the horizon and the end of inflation.

The scale-dependence of the perturbations is described by the
spectral tilt \cite{Maartens}

\beq 
n_{s} - 1 \equiv {d\ln A_{s}^2 \over d\ln k} \simeq -6\epsilon + 2\eta~~,
\label{eq:ns}
\eeq
 which, for this model, gives

\begin{equation}
n_{s} = 1-192{\beta^2\over{\alpha^2}}
\exp\left({8 N_k\over{\alpha}}\right) + {8\over{\alpha}}~.
\label{eq:ns2}
\end{equation}

The ``running'' of the scalar spectral index is given by

\beq
\alpha_s\equiv {d n_s \over d\ln k}
=16\epsilon\eta-24\epsilon^2-2\xi~,
\label{eq:alfas}
\eeq
and we get 

\begin{equation}
\alpha_{s} = 512{\beta^2\over{\alpha^2}}\exp\left({8 N_k\over{\alpha}}\right)
\left[1-3\exp\left({8 N_k\over{\alpha}}\right)\right]~.
\end{equation}

The amplitude of tensor  perturbations is given by \cite{Langlois}

\beq
A_t^2=\left.{64\over 150\pi M_P^4} V\left( 1+{V\over 2 \lambda}\right) F^2
 \right|_{k=aH}~,
\eeq
where

\beq
F^2=\left[\sqrt{1+s^2} -s^2 \sinh^{-1}\left({1\over s}\right)\right]^{-1}~,
\eeq
and
\beq
s\equiv \left[{2V\over \lambda }\left(1+{V\over 2\lambda}\right)\right]^{1/2}~.
\eeq
In the low energy limit ($s\ll 1$), $F^2\approx 1$, whereas
$F^2\approx 3 V/2\lambda$
 in the high energy limit. Defining (we choose the normalization of
 Ref.~\cite{Peiris})

\beq
r_s\equiv 16 {A_t^2\over A_s^2}~,
\eeq

we obtain

\beq
r_s\simeq 0.06 {M_P^4 m^4\lambda\beta^2\over V_0^3}
\exp\left(8 N_k\over \alpha\right)~.
\eeq

\begin{figure}[ht!]
\centering
\leavevmode \epsfysize=7.5cm \epsfbox{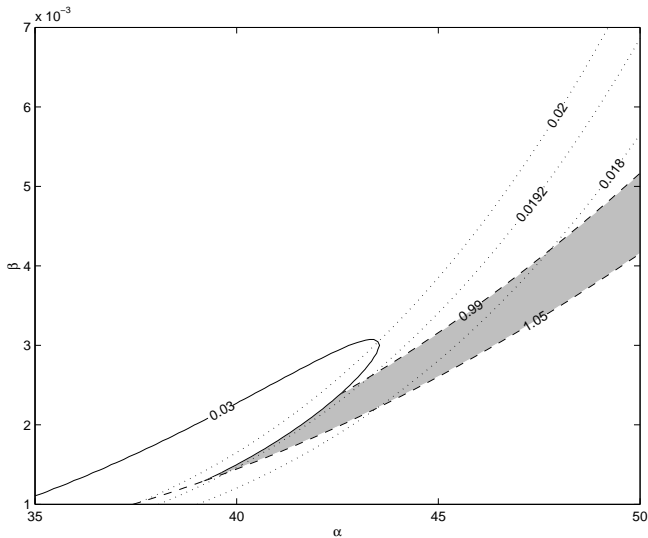}
\vskip 0.1cm
\caption{Contours of the inflationary observables  $n_s$ (dashed) and
  $\alpha_s$ (full) for $N_\star=55$.  The allowed region is shaded
  and the dotted   contours correspond 
to different values of $M_5$ in units of $M$.}
\label{fig:figure2}
\end{figure}

WMAP bounds on the above inflationary observables are, for this class
 of models (case $\eta>3\epsilon$, class D in Ref.  \cite{Peiris})
\beqa
&& 0.99 <   n_s < 1.28~, \quad -0.09\leq \alpha_s\leq 0.03~,\nonumber\\
 & & r_s \leq 1.10 ~.
\label{eq:observations2}
\eeqa
These bounds refer to the scale best probed by the CMB observations
i.e. $k=0.002$ Mpc$^{-1}$; accordingly, we set $N_k(k=0.002)=N_\star$.
 On the other hand, bounds on $n_s$ from
other experiments are less blue, e.g. the combined data sets from
BOOMERANG, CBI, DASI, DMR, MAXIMA, TOCO and VSA give \cite{comb}:

\beq
 0.955< n_s < 1.05~.
\label{eq:comb}
\eeq

In Figure \ref{fig:figure1}, we show contours
of the observational bounds on the inflationary observables 
$n_s$, $r_s$ and $\alpha_s$ in the
 ($\alpha$, $\beta$) parameter space, for different values of
 $N_\star$; in these plots, we have taken the bounds of
 Eq.~(\ref{eq:observations2})
except for the upper bound on $n_s$, for which we took the  bound of
Eq.~ (\ref{eq:comb}) instead since such a blue spectrum is not to be expected.
  We also plotted
 the
 $\alpha_s=0$ contour, which shows
that $\alpha_s$ is required to be positive for this model. The shaded area
 corresponds to the allowed region 
in parameter space. 
We would also like to mention that we have checked that it is possible to
 obtain sufficient inflation 
with sub-Planckian field values e.g $N=70$ for $\phi_I=0.2 M_P$ and
 $\alpha,~\beta$ within the range specified in Figure 1. 

\begin{figure}[ht!]
\centering
\leavevmode \epsfysize=7cm \epsfbox{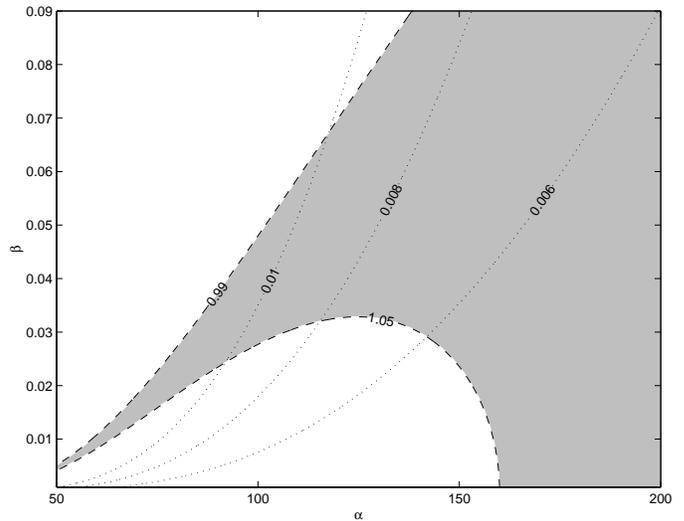}
\vskip 0.1cm
\caption{Contours of the inflationary observable  $n_s$ (dashed) 
 for $N_\star=55$ (notice that the range of
  $\alpha$ and $\beta$ is enlarged as compared with Figure
  \ref{fig:figure2}). The allowed region is shaded and 
 the dotted  
 contours correspond 
to different values of $M_5$ in units of $M$. }
\label{fig:figure3}
\end{figure}
 
 Notice that  the contour corresponding to the upper bound
on $\alpha_s$ is too small to be visible in Figure \ref{fig:figure1},
hence 
we show it in
Figure \ref{fig:figure2}, for $N_\star=55$ (similar behavior is
obtained for other values of  $N_\star$), which makes it  clear that
 this bound plays
 an important role in constraining the parameter space. We obtain
 lower bounds on $\alpha$ and $\beta$, namely  $\alpha > 39.23$,
 pratically independent of $N_\star$, whereas  the lower bound on
 $\beta$ ranges from $2.2\times 10^{-3}$ to $1.7\times 10^{-4}$ as
 $N_\star$ ranges from 50 to 75.

In Figures \ref{fig:figure2} and \ref{fig:figure3}, we have superposed
contours of the scale $M_5$, as derived from Eqs.~(\ref{eq:As2}) and
(\ref{eq:MP}), where we have used the COBE normalization i.e.
$A_s=2\times 10^{-5}$ for $N_k=N_\star=55 $.  As the allowed region is
quite narrow for low values of $\alpha$, it allows us to find
 an  upper
bound on $M_5$, $M_5 \lsim 0.0194$ M, which  we have checked is almost
independent of $N_\star$. Combining the above results, we find a lower
bound
 on the scale $V_0$,
namely
$V_0^{1/4} \lsim 2.1 \times 10^{-3}$ M.

Notice that we have chosen to vary $N_\star$ since, although
a wide variety of assumptions about $N_\star$ can be found in the
literature, the determination of this quantity requires a model of the
entire history of the Universe. However, while from nucleosynthesis
onwards this is now well established, at earlier epochs there are
considerable uncertanties such as the
mechanism ending inflation and  details of the reheating process.
This issue was recently reviewed in Ref.~\cite {Liddle2} (see also 
Ref.~\cite{Dodelson} for similar results), where a model-independent
 upper bound was derived,
namely $N_\star<60$; in fact, $N_\star=55$
is found to be a reasonable fiducial value with an uncertainty of
around 5 around that value; however, the authors stress that there
 are several ways in which $N_\star$ could
lie outside that range, in either direction. Moreover, in the
braneworld context, one expects $N_\star$ to depend on the brane
tension. Actually,  one expects to obtain larger
values of $N_\star$  because, in the high-energy regime,
the expansion laws corresponding to matter and radiation domination
are slower than in the standard cosmology, which implies a greater
change in $a H$ relative to the change in  $a$, therefore  requiring a
 larger value of
$N_\star$. This is confirmed by the results of  Ref. \cite{Wang},
 where the  bound
 $N_\star < 75$ is found for brane inspired cosmology.

We have studied the dependence of $n_s$ on $k$ to see whether $n_s$
can vary from $n_s>1$ to $n_s<1$ from large to small scales. From the
condition $k=aH$, assuming that H is approximatly constant during
inflation, we obtain the relation 

\beq
\exp(N_k)={k_F\over k}~,
\label{eq:Nkvk}
\eeq
where $k_F$ is the value of $k$ at the end of inflation. Inserting
this relation is Eq.~(\ref{eq:ns2}), we obtain

\beq
n_s\approx 1+{8\over \alpha}-192 {\beta^2\over \alpha^2}\left({k_F\over
  k}\right)^{8/\alpha}~,
\label{eq:nsk}
\eeq
from which we conclude that, for $\alpha,~\beta$ fixed, $n_s$
increases with $k$. Hence, it is not possible to obtain the desired
behaviour, i.e. $n_s$ decreasing from $n_s>1$ to $n_s<1$ as $k$ increases.

\begin{figure*}[ht!]
\centering
\leavevmode \epsfysize=14cm \epsfbox{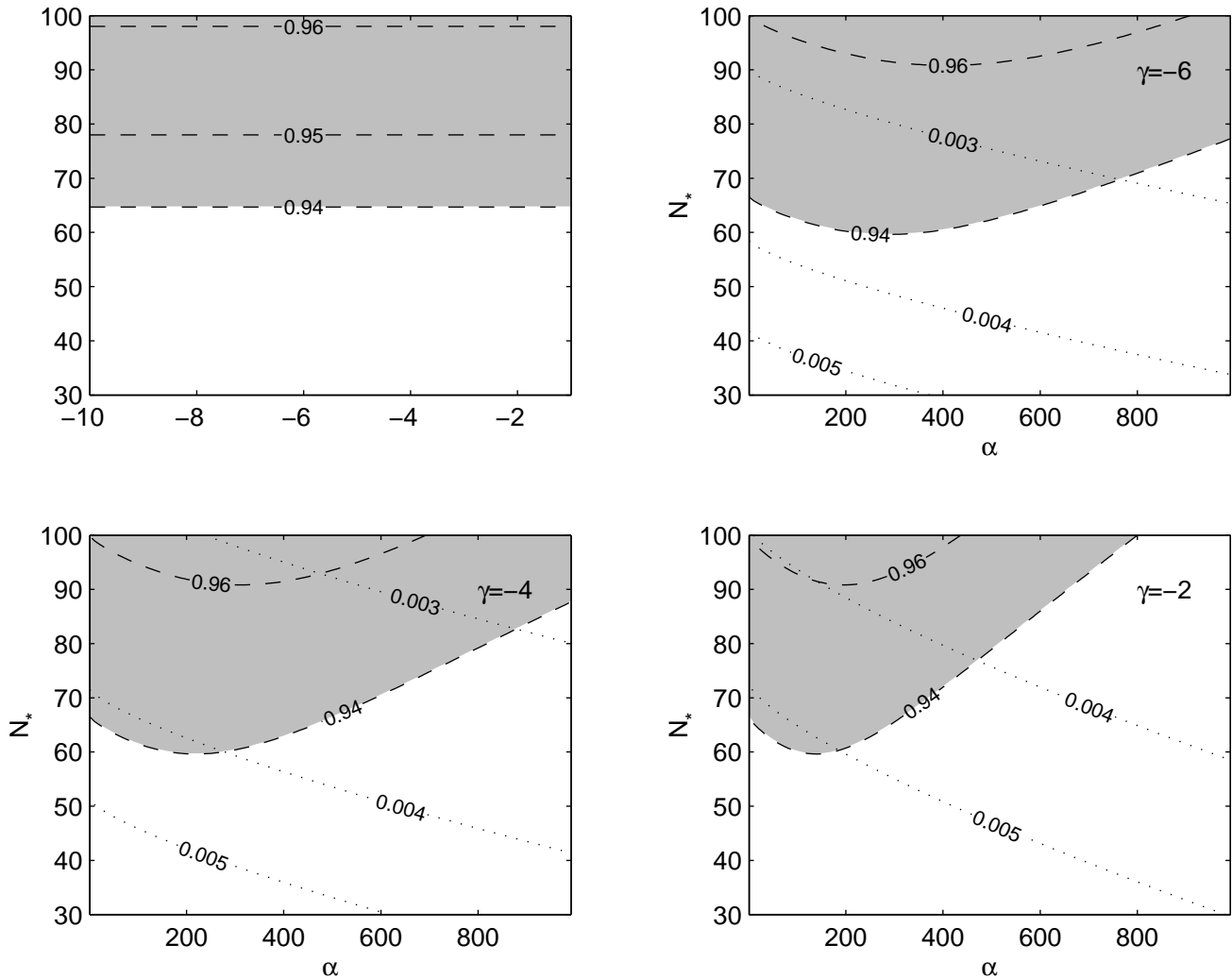}\\
\vskip 0.1cm
\caption{Contours of the inflationary observables  $n_s$ (dashed) 
 in the ($\alpha, N_\star$) plane for
 the cubic model. The allowed region is shaded and  
 we also show  contours corresponding 
to different values of $M_5$ (dotted) in units of $M$. The upper left
panel
 corresponds to the results for this model in the low energy approximation,
 i.e. without the brane corrections,
 and the remaining three panels correspond to the high energy approximation 
  for  different values of $\gamma$.}
\label{fig:figure4}
\end{figure*}
%
\section{Cubic potential}

We shall now consider the case where, due to some cancellation 
mechanism \cite{Adams}, the quadratic term is absent and the potential
 is cubic in 
$\phi$:

\beq
V=\Delta^4\left[1 + \gamma \left(\phi\over M\right)^3\right]~~.
\label{eq:V3}
\eeq
As mentioned before, we shall assume that the first term is dominant.
The parameter $\gamma $ is expected to be of order unity and
negative \cite{Adams};
the model of Ref.~\cite{Ross} corresponds to precisely 
this case, with $\gamma=-4$.

We start by computing the slow-roll parameters:
\beqa
\epsilon &\simeq &{18 \gamma^2 \over \alpha} \left({\phi \over
  M}\right)^4~,\nonumber\\ 
\eta &\simeq& {12 \gamma \over \alpha} \left({\phi \over M}  \right)~,
\nonumber\\
\xi& \simeq & {72 \gamma\over \alpha^2} \phi^2~,
\label{eq:slowr3}
\eeqa
where $\alpha\equiv {\Delta^4\over \lambda}$.

The value of $\phi$ at the end of inflation can be obtained from 
Eq.~(\ref{eq:phif}); we get, from $|\epsilon| \simeq 1$ 

\beq
\phi_F \simeq \left({\alpha \over 18 \gamma^2}\right)^{1/4} M~,
\label{eq:phif3}
\eeq
while, from $|\eta| \simeq 1$, we obtain
\beq
\phi_F \simeq \left({\alpha \over 12 |\gamma|}\right) M~.
\label{eq:phif4}
\eeq
Hence, the prescription to be used depends on the value of $\alpha$. 
For $\gamma = - 4$, we see that the two prescriptions coincide 
for $\alpha \simeq 26$. 

The number of $e$-foldings, $N$, is given by:

\beq
N  =  {\alpha M \over 6 |\gamma|} \left[{1 \over \phi_I} - 
{1 \over \phi_F}\right]~~.
\label{eq:N3}
\eeq  
Therefore, sufficient inflation
to solve the cosmological horizon/flatness problems, that is $N > 70$, 
is achieved,  for instance for  $\gamma=-4$, if
$\phi_I < 7.5 \times 10^{-2} M_P$.

For $A_s$, we obtain, in the high energy regime,

\beq
A_s^2 \simeq  {\alpha^4 \lambda \over 5400~\pi^2 \gamma^2  \phi_k^4}~~,
\label{eq:As3}
\eeq
where $\phi_k$ is the value of $\phi$ at horizon-crossing. 
The scalar spectral index and its running can be readily computed from
the slow-roll parameters, Eq.~(\ref{eq:slowr3}), via
Eqs.(\ref{eq:ns}) and   (\ref{eq:alfas}). Notice that the  inflationary 
observables can, of course,  be written as a function
of $N_\star$, as for the quadratic model,  using Eq.~(\ref{eq:N3})
 with $\phi_I=\phi_k$, but one
has to bear in mind that the prescription to use for $\phi_F$ depends
on $\alpha$.

WMAP bounds on the inflationary observables are, for this class of models 
 (case $\eta<0$, class A in Ref.~\cite{Peiris})
 \beqa
&& 0.94 <  n_s < 1.00~, \quad -0.02\leq \alpha_s\leq 0.02~,\nonumber\\
&&  r_s \leq 0.14 ~,
\label{eq:observations}
\eeqa
again for the scale $k=0.002$ Mpc$^{-1}$.
 In Figure \ref{fig:figure4}, we show contours of the inflationary
 observable  $n_s$, $\alpha_s$ in the ($\alpha, N_\star$) plane. We
 have checked that neither $\alpha_s$ nor $r_s$ give further
 constraints on the parameter space.  
 We also show  contours corresponding 
to different values of $M_5$, as given by Eq.~(\ref{eq:As3}), again
 COBE normalized.
 The upper left panel   corresponds  to the results for the low
 energy regime, 
$V\ll \lambda$, where the brane corrections are negligible, and the
 remaining three panels correspond to the high energy regime, 
$V\gg \lambda$, for  different values of $\gamma$.

We see that it is the lower bound on $n_s$ that most constrains the
model and, clearly, $n_s=1$ cannot be obtained.
It is also clear that, for the model to work, $N_\star>65$ is
 required if brane corrections are not included and $N_\star>60$ if
 those  corrections are included; in the latter case, however, this
 bound increases outside the range $300 \lsim \alpha \lsim 100$, for
 $\gamma=-4$
 (this range is slightly  $\gamma$-dependent, see Figure
 \ref{fig:figure4}).
Moreover, the running parameter is always negative although it can be
quite small. Finally,
 $0.042\ M \lsim M_5 \lsim 0.025\ M$,  for $\gamma=-4$ (however, these
bounds do not change
significantly with $\gamma$, see Figure 4. 

Clearly, the spectral index cannot run from
$n_s>1$ on large scales to 
 $n_s<1$ on small scales, since $n_s(k)<1$
 for this model.

In Ref.~\cite{Bento2}, a very strict bound on $\alpha$ was derived for
this model from the requirement that the reheating temperature is
small enough to avoid the gravitino problem. We should like to point
out that there was a numerical error in that computation and, in fact,
the bound is much weaker and pratically meaningless.

\section{Conclusions}

We have analysed the implications of WMAP results, in particular the
bounds on the inflationary observables, for a class of 
supergravity inflationary models, Eq.~(\ref{eq:gpot}) with $n=2,~3$.
We find that, for the quadratic potential, the main constraints come from  the
WMAP's  bounds on $n_s$ and upper bound on $\alpha_s$. 
We have obtained lower  bounds on parameters $\alpha$ and $\beta $,
 namely $  \alpha \gsim 39$ (pratically independent of $N_\star$) and 
the lower bound on  $\beta$ ranges from $2.2 \times 10^{-3}$ to
 $1.7\times 10^{-4}$  as  $N_\star$ varies
between 50 and 75.
We have also found
 an  upper
bound on $M_5$, $M_5 \lsim 0.0194$ M, pratically
independent of $N_\star$. Moreover, we conclude that $\alpha_s>0 $ is
required for this model.

For the cubic potential, in the low energy regime i.e. without the
brane 
correction, a relatively high value of $N_\star$, $N_\star >65$,
 is required so as to meet WMAP's lower bound
on $n_s$.
In the high energy regime, when brane corrections are significant, the
allowed region in the ($\alpha, N_\star$) parameter space changes
with $\gamma$ and the main constraints come from WMAP's lower bound on
$n_s$. Moreover, we find that
$\alpha_s<0$ for this model

We have also studied whether it is possible to obtain a running
spectral index such that $n_s>1$ on large scales and $n_s<1$ on
 small scales and concluded that this is not possible for either
 model.

\vfill
\begin{acknowledgments}

\noindent
M.C.B. acknowledges the partial support of Funda\c c\~ao para a 
Ci\^encia e a Tecnologia (Portugal)
under the grant POCTI/1999/FIS/36285. The work of A.A. Sen is fully 
financed by the same grant.  N.M.C. Santos is supported by FCT grant
SFRH/BD/4797/2001.

\end{acknowledgments}


\begin{thebibliography}{99}

\bibitem{Bennet} C.L. Bennet {et al.}, \AJ {\it Suppl.}  {\bf
  148},  (2003) 1.

\bibitem{Spergel} D.N. Spergel {et al.}, \AJ {\it Suppl.}  {\bf
 148} (2003) 175.

\bibitem{BargerSeljak} V. Barger, H. Lee and D. Marfatia,
 hep-ph/0302150;  U. Seljak, P. McDonald, A. Makarov, \MNRAS
{\bf 342} (2003) L79. 

\bibitem{Bertolami} O. Bertolami, G.G. Ross, \PL {\bf B183} (1987) 163.

\bibitem{Randall} L. Randall and R. Sundrum, \PRL {\bf 83}
 (1999) 4690.

\bibitem{Binetruy} P. Bin\'etruy, C. Deffayet, U. Ellwanger, D. Langlois,
\PL {\bf B477} (2000) 285;

T. Shiromizu, K. Maeda, M. Sasaki, \PR {\bf D62} (2000) 024012.

E.E. Flanagan, S.H. Tye, I. Wasserman, \PR {\bf D62} (2000) 044039.

\bibitem{Bento2}  M.C. Bento,
 O. Bertolami,  A.A. Sen,  \PR {\bf D67} (2003) 023504.

\bibitem{Maartens} R. Maartens, D. Wands, B.A. Bassett, I.P.C. Heard,
\PR {\bf D62} (2000) 041301.


\bibitem{Bento1} M.C. Bento, O. Bertolami, \PR {\bf D65} (2002)
  063513.

\bibitem{Liddle1}  A.R. Liddle, A.J. Smith, \PR {\bf D68}
 (2003) 061301.

\bibitem{Langlois} D. Langlois, R. Maartens, D. Wands, \PL {\bf B489}
(2000) 259.

\bibitem{Peiris}  H.V. Peiris {et al.}, \AJ {\it Suppl.}  {\bf
 148} (2003) 213.

\bibitem{comb} J.L. Sievers et al., \AJ {\bf 591}
  (2003) 599.

\bibitem{Liddle2} A.R. Liddle, S.M. Leach, \PR {\bf 68} (2003) 103503.


\bibitem{Dodelson} S. Dodelson, L. Hui, \PRL  {\bf 91}
  (2003) 131301.

\bibitem{Wang} B. Wang, E. Abdalla, hep-th/0308145.

\bibitem{Adams} J.A. Adams, G.G. Ross, S. Sarkar, \PL {\bf B391}
  (1997) 271.

\bibitem{Ross} G.G. Ross, S. Sarkar, \NP {\bf B461} (1995) 597.



\end{thebibliography}
\end{document}